\documentclass[12pt,preprint]{aastex}
\shorttitle{Galaxy Spins in cDE Models}
\begin{document}
\title{Detection of the Vorticity Effect on the Disk Orientations}
\author{Jounghun Lee}
\affil{Department of Physics and Astronomy, Seoul National University, Seoul 
151-747, Korea}
\email{jounghun@astro.snu.ac.kr}
\begin{abstract}
We present an observational evidence for the vorticity effect on the nonlinear evolution of the 
galaxy angular momentum. 
We first calculate the vorticity as the curl of the peculiar velocity 
field reconstructed from the 2MASS redshift survey. Then, measuring the 
alignments between the vorticity and the tidal shear fields, we study how the alignment 
strength and tendency depends on the cosmic web environment. It is found that in the 
knot and filament regions the vorticity vectors are anti-aligned with the directions of the maximal 
volume compression while in the void and sheet regions they are anti-aligned with the directions 
of the minimal compression. 
Determining the spin axes of the nearby large face-on and edge-on disk galaxies from the Seventh 
Data Release of the Sloan Digital Sky Survey and measuring their correlations with 
the vorticity vectors at the galaxy positions, we finally detect a clear signal of the alignments 
between the galaxy spin and the local vorticity fields. The null hypothesis that there is no 
alignment between them is rejected at the $99.999\%$ confidence level.
Our result supports observationally the recently proposed scenario that although the 
galaxy angular momentum originates from the initial tidal interaction in the linear 
regime its subsequent evolution is driven primarily by the cosmic vorticity field.
\end{abstract}
\keywords{cosmology:theory --- large-scale structure of universe}
\section{INTRODUCTION}\label{sec:intro}

The most prevalent theory for the origin of the galaxy spins is the tidal torque theory which explains 
that a galaxy acquired its spinning motion (i.e., the angular momentum) via the tidal interaction with the 
surrounding matter fluctuations at its proto-galactic stage \citep{peebles69,dor70,white84}. 
One of the key predictions of the tidal torque theory is the existence of the local correlations between 
the galaxy spin and the  tidal shear fields 
\citep{CT96a,CT96b,LP00,PLS00,LP01,porciani-etal02a,porciani-etal02b}: 
The initial tidal torques which lasted till the turn-around moments caused the alignments between the spin axes 
of the proto-galaxies and the intermediate principal axes of the linear tidal fields. In the subsequent 
evolution the initially established spin-shear alignments would become weaker as the non-linear effects like 
galaxy-galaxy interactions should play a role of randomizing the galaxy spin axes 
\citep{dubinski92,CT96b,cri-etal01,cri-etal02,porciani-etal02a,LP02,LP08}. 

Plenty of observational evidences have been found for the existence of the intrinsic spin-shear correlations
 \citep[][for a comprehensive review]{schafer09}. 
For instance, \citet{LE07} provided a direct observational evidence for the existence of the intrinsic
spin-shear correlations by using the tidal shear fields reconstructed from the 2MASS 
redshift survey \citep{erdogdu-etal06} and the galaxy catalog complied by B. Tully \citep{PLS00}. 
However, even though the main result of \citet{LE07} was seemingly consistent with the prediction of the linear 
tidal torque theory, they noted that  the detected environmental dependence of the spin-shear alignments could 
not fit well into the tidal torque picture. The nonlinear effects which would destroy the initially established spin-
shear alignments should be stronger in denser environments.  Therefore, in the tidal torque picture, the galaxies 
located in denser regions were expected to show weaker spin-shear alignments.  In contrast,  what was found 
by \citet{LE07} was an opposite trend that the galaxies located in denser environments exhibit stronger intrinsic 
spin-shear alignments \citep[see Figure 6 in][]{LE07}.  

A similar trend was also detected by \citet{lee11} who measured the spin-spin alignments by using the nearby 
disk galaxies from the seventh data release of the Sloan Digital Sky Survey (SDSS DR7) \citep{sdssdr7}. 
They showed that the galaxies which have ten or more neighbors within separation distance of 
$2\,h^{-1}$Mpc exhibit stronger spin-spin alignments.  If the intrinsic spin-spin alignments were related to the 
tidally induced spin-shear alignments  
\citep[e.g.,][]{PLS00,cri-etal01,cri-etal02,jing02,mac-etal02,porciani-etal02a,porciani-etal02b}, 
this observational finding of \citet{lee11}  should be interpreted as another evidence for the stronger 
spin-shear alignments in denser environments, consistent with the result of \citet{LE07}.  
Although \citet{lee11} attempted to explain the observed environmental dependence of the intrinsic spin-spin 
alignments in the tidal torque picture, ascribing it to the dependence of the tidal strength on the local density,  it 
remains unanswered why and how the initially established spin-shear alignments survive better in denser 
environments. 

Several literatures have also suggested that the tidal torque theory alone could not provide a fully satisfactory 
explanation for the observed dependence of the galaxy intrinsic alignments on the cosmic web environment 
\citep[see, e.g.,][]{hah-etal07b,jones-etal10,codis-etal12,libes-etal12a,GS13,tro-etal13}.
The general consensus reached by the previous numerical works is that the observed galaxy intrinsic 
alignments are not just weakened version of the initial spin-shear correlations but evolved version driven 
by some other mechanism than the tidal shears. What has so far suggested as a possible candidate 
for the required mechanism in the literatures includes the large-scale cosmic flow, peculiar motions, satellite 
infalls, bulk flows and etc. 
 
Very recently, \citet{libes-etal12b} proposed an interesting new scenario that the cosmic vorticity drives the 
nonlinear evolution of the galaxy angular momentum after the turn-around moments.  In the linear regime the 
peculiar velocity field is curl free as it is described as a gradient of the perturbation potential. 
In the subsequent nonlinear evolution, however, it develops the curl mode which could affect the galaxy angular 
momentum.  By analyzing the data from high-resolution N-body simulations, \citet{libes-etal12b} 
demonstrated that the halo angular momentum vectors are strongly aligned with the local vorticity vectors, 
indicating the presence of the dominant effect of the vorticity field on the nonlinear evolution of the 
galaxy angular momentum vectors.
The \citet{libes-etal12b} also showed that the vorticity vectors are strongly anti-aligned with the major 
(minor) principal axes of the local tidal shear fields in the knot (void) environments. 

We note here that the puzzling environmental dependence of the galaxy spin alignments can be coherently 
explained by assuming the vorticity driven evolution of the galaxy angular momentum. The goal of this 
Paper is to look for an observational evidence for this new scenario of \citet{libes-etal12b}.  To achieve this goal, 
we reconstruct the cosmic vorticity and the galaxy spin fields from the large galaxy surveys and investigate if the 
vorticity-spin and vorticity-shear alignments found in N-body simulations  exist in the real Universe. The 
upcoming three sections present the following highlights, respectively: the procedure to reconstruct the cosmic 
vorticity field and the result on the vorticity-shear alignments in section \ref{sec:vor}; a brief review of the 
algorithm to measure the galaxy spin axes with high accuracy and the result on the vorticity-spin alignments in 
section \ref{sec:spin}; a concise summary of the final results and discussion of the cosmological 
implication of our results in section \ref{sec:conclusion}.

\section{VORTICITY-SHEAR ALIGNMENTS}\label{sec:vor}

\citet{erdogdu-etal06} reconstructed the density  contrast $\delta({\bf x})$ and peculiar velocity ${\bf v}({\bf x})$ 
fields by applying the Wiener reconstruction algorithm  to the galaxy data from the 2MASS redshift survey 
\citep{2mass} under the assumption of a $\Lambda$CDM universe. The fields were provided on $64^{3}$ 
pixels in a regular cubic space of  linear size $400\,h^{-1}$Mpc, for which the position vectors ${\bf x}$ were 
expressed in the supergalactic coordinate system and the peculiar velocities ${\bf v}$ were measured 
in the cosmic microwave background (CMB) frame.  Given the spatial resolution of $ 6.25\,h^{-1}$Mpc, 
the reconstructed fields represented  the density and velocity fluctuations on the quasi-nonlinear scale.  
 \citet{erdogdu-etal06} proved the robustness of their reconstruction procedure by demonstrating that 
the predicted density field recovers well the observed large scale structures of  the Universe. 

Using the density field $\delta({\bf x})$ reconstructed by \citet{erdogdu-etal06} from the 2MASS redshift survey, 
\citet{LE07} calculated the tidal fields as $T_{ij}({\bf x})=\partial_{i}\partial_{j}\nabla^{-1}\delta({\bf x})$.  
Given that on those pixel points more distant than $100\,h^{-1}$Mpc from the center  the reconstructed 
density and velocity fields suffered from large uncertainties (private communication with P.Erdogdu),
 \citet{erdogdu-etal06} selected only those $32^{3}$ pixels whose separation distances from the center  
are less than $100\,h^{-1}$ Mpc and determined the major, intermediate and minor principal axes of 
$T_{ij}({\bf x})$ at each selected pixel point. The detailed descriptions of  the reconstructed density, 
velocity and tidal fields, and the 2MASS redshift survey can be found in \citet{erdogdu-etal06}, 
\citet{LE07} and \citet{2mass}, respectively. 

\begin{figure}[ht]
\begin{center}
\plotone{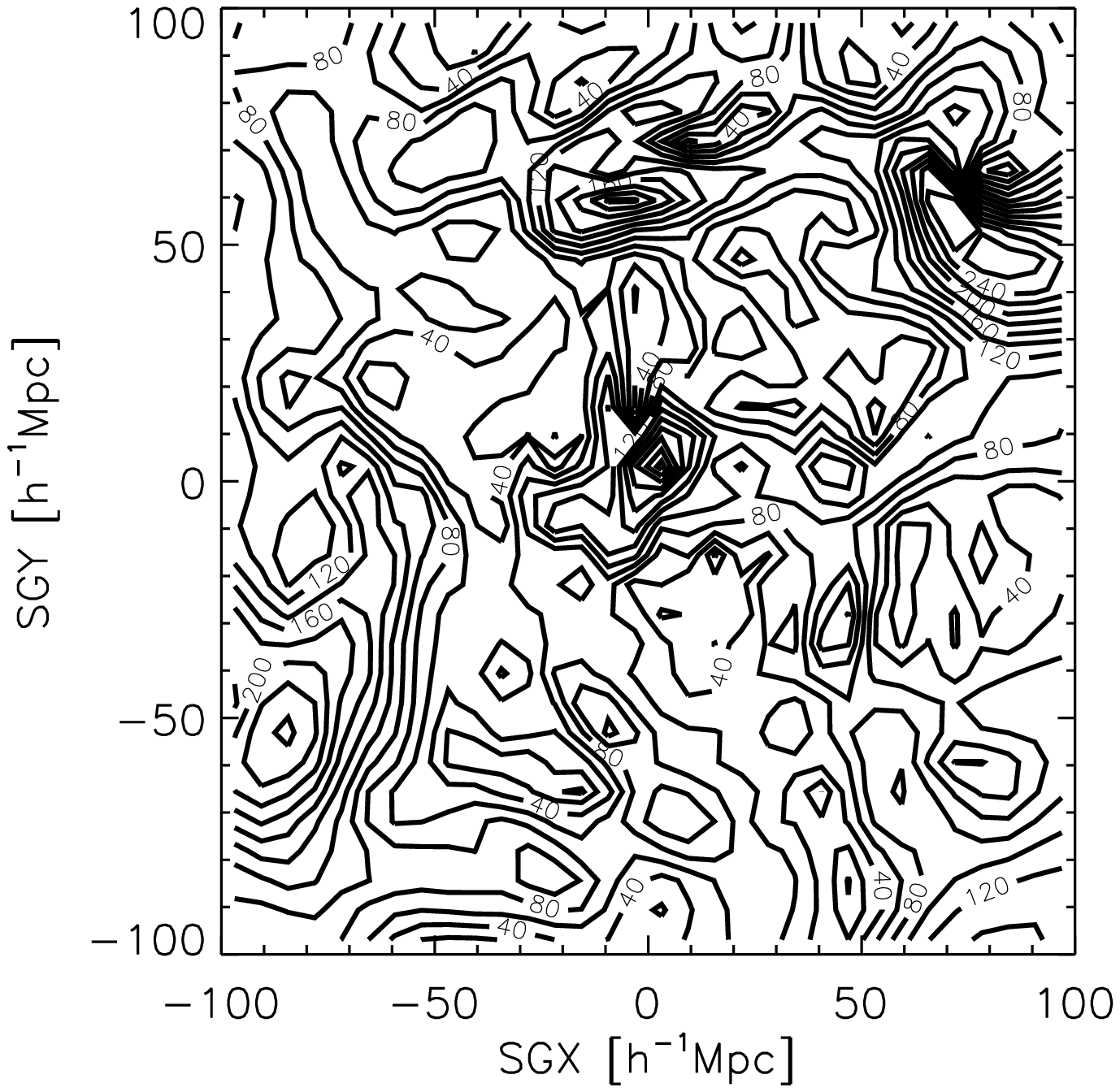}
\caption{Contours of the magnitudes of the real space vorticity fields 
reconstructed from the 2MASS redshift survey in the supergalactic 
$x$-$y$ plane.}
\label{fig:contour}
\end{center}
\end{figure}

Now, we attempt to reconstruct the cosmic vorticity field on the selected $32^{3}$ pixel points by calculating the 
curl of the peculiar velocity field as ${\bf w}\equiv{\bf \nabla}\times{\bf v}$.  We first perform the Fourier transform 
of the peculiar velocity field to obtain its Fourier amplitudes $\tilde{\bf v}$ with the help of the Fast Fourier 
Transformation (FFT) code \citep{pre-etal92}. The Fourier amplitude of the vorticity field 
can be written as $\tilde{\bf w}={\bf k}\times\tilde{\bf v}$ where ${\bf k}$ is the wave vector in the 
Fourier space, and thus the three components of $\tilde{\bf w}$ are calculated in Fourier space as 
\begin{equation}
\tilde{w}_{1}=k_{2}\tilde{v}_{3}-k_{3}\tilde{v}_{2}\, ,\qquad
\tilde{w}_{2}=k_{3}\tilde{v}_{1}-k_{1}\tilde{v}_{3}\, ,\qquad
\tilde{w}_{3}=k_{1}\tilde{v}_{2}-k_{2}\tilde{v}_{1}\, .
\end{equation}
Finally, we perform the inverse Fourier transform of $\tilde{\bf w}$ to obtain the real-space 
vorticity field ${\bf w}$.  Figure \ref{fig:contour} plots the contours of the absolute magnitude of 
${\bf w}$ in the supergalactic $x$-$y$ plane,  showing the deviation of $\vert{\bf w}\vert$ from zero. 
Recalling the fact that the peculiar velocity field reconstructed from the 2MASS redshift survey 
was filtered on the quasi-nonlinear scale of $\sim 6.25\,h^{-1}$Mpc, the result shown in Figure 
\ref{fig:contour} implies that  even in the quasi-nonlinear regime the velocity field is no longer irrotational,  
developing the curl mode.

As done in \citet{libes-etal12b}, we first divide the $32^{3}$ pixels into the knots, filaments, sheets and voids 
according to the signs of the shear eigenvalues \citep{hah-etal07a}: The pixels at which  all three eigenvalues 
have positive (negative) values are marked as knots (voids), while the pixels at which one eigenvalue is 
negative (positive) and the other two are positive (negative) are marked as filaments (sheets).  Then, we 
calculate the alignment between the vorticity vector and the principal axes of the tidal tensor at  each marked 
pixel. Let $\{{\bf e}_{1},\  {\bf e}_{2},\ {\bf e}_{3}\}$ be the major, intermediate and minor principal axes of the tidal 
tensor at each marked pixel. 
Calculating $\mu\equiv \vert{\bf w}\cdot{\bf e}_{i}\vert $ (for $i=1,\ 2,\ 3$) and binning the values of $\mu$ 
in the range of $[0,\ 1]$,  we determine the probability density distribution of $\mu$.  If there were no alignment, 
the distribution $p(\mu)$ would be uniformly unity. If $p(\mu)$ increases (decreases) with $\mu$, then there 
should be strong alignments (anti-alignments) between the vorticity vectors and the principal axes of the 
tidal tensors. We repeat the same calculation for the knot, void, filament and sheet regions to separately 
determine $p(\mu)$ for each case.

\begin{figure}[ht]
\begin{center}
\plotone{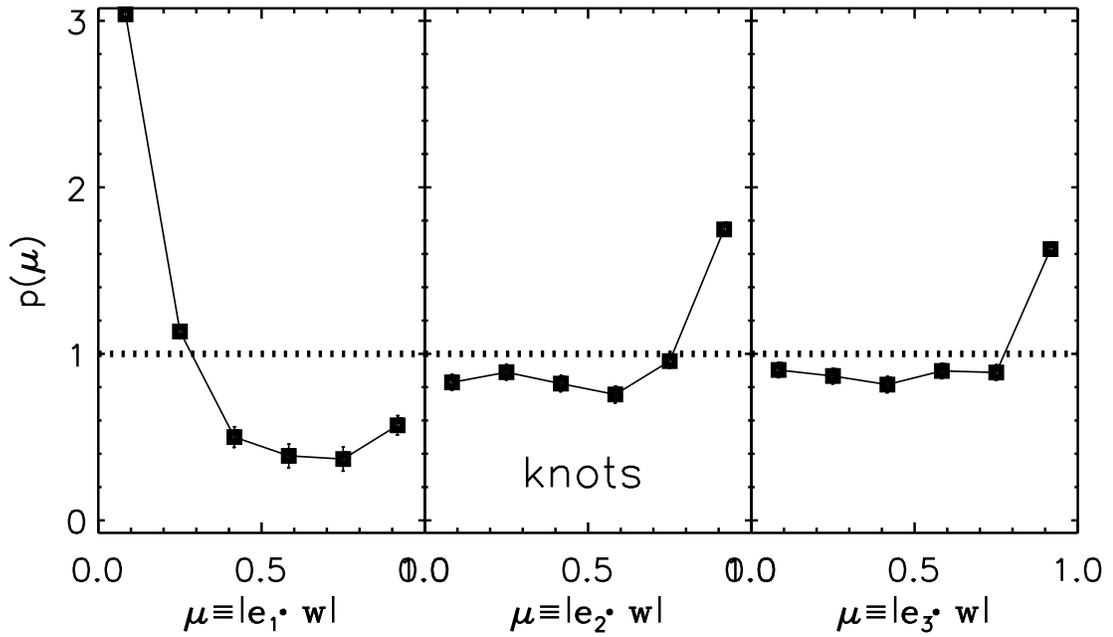}
\caption{Probability density distributions of the cosines of the angles between 
the vorticity vectors and the major, intermediate and minor principal axes of the 
tidal shear tensors in the left, middle and right panels, respectively, for the knot regions 
where the eigenvalues of the tidal shear tensors are all positive. 
In each panel, the dotted line corresponds to the case of no alignment and the errors are Poissonian.}
\label{fig:evor_knot}
\end{center}
\end{figure}

Figure \ref{fig:evor_knot} plots the probability density distributions , $p(\mu)$, wit Poisson errors for the knot 
regions, showing how the vorticity vectors are aligned with the major, intermediate and minor principal axes of 
the tidal tensors in the left, middle and right panels, respectively.  In each panel the horizontal dotted line 
indicates the uniform distribution of $\mu$ for the case of no vorticity-shear alignments. 
As can be seen,  the vorticity vectors  are strongly anti-aligned with the major principal axes of the tidal 
shear tensors, preferentially lying in the plane spanned by their intermediate and minor principal axes in the knot 
regions. In other words, in the highly dense knot environments the vorticity vectors are anti-aligned with the 
directions of the maximum matter compression. 

\begin{figure}
\begin{center}
\plotone{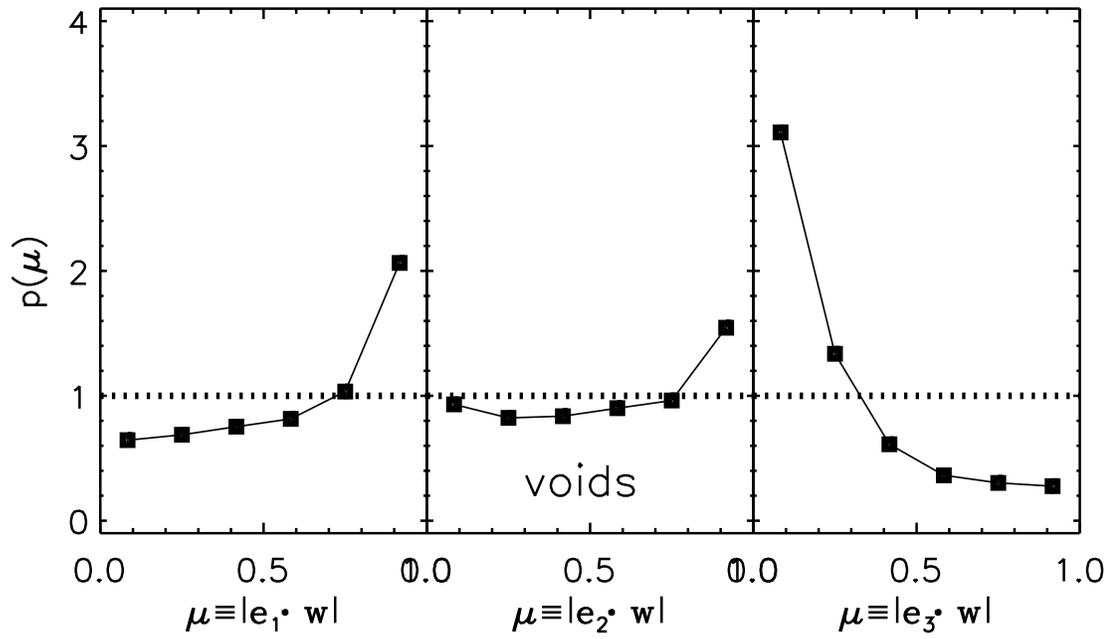}
\caption{Same as Figure \ref{fig:evor_knot} but for the void regions where the shear 
eigenvalues are all negative.}
\label{fig:evor_void}
\end{center}
\end{figure}

\begin{figure}
\begin{center}
\plotone{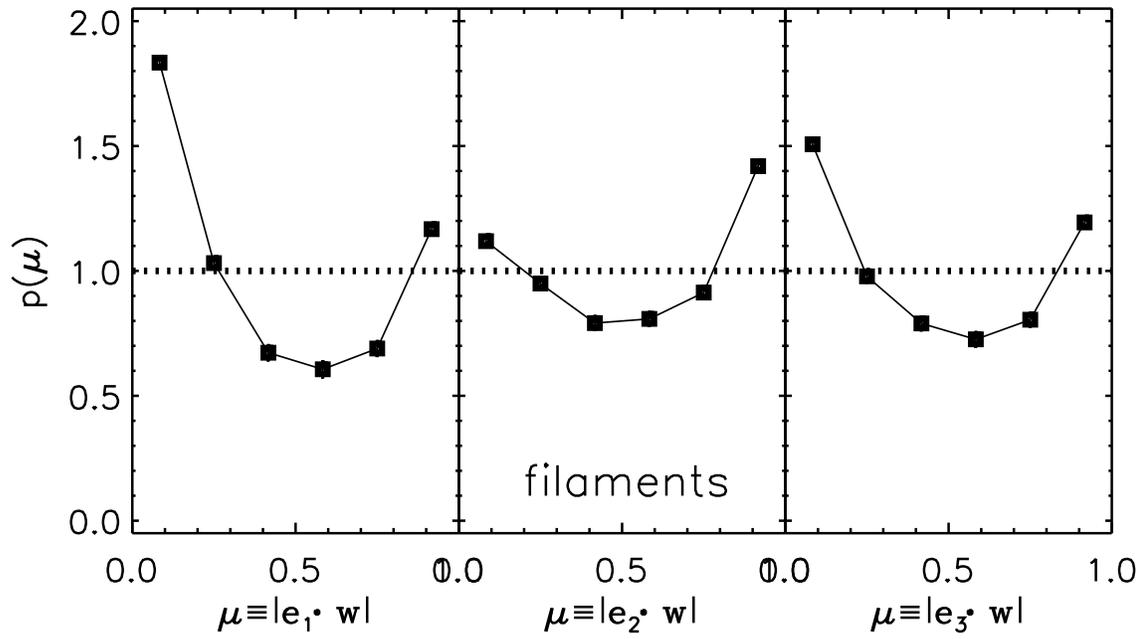}
\caption{Same as Figure \ref{fig:evor_knot} but for the filament regions that the 
largest and second to the largest shear eigenvalues are positive while the smallest eigenvalues 
are negative.}
\label{fig:evor_pan}
\end{center}
\end{figure}

\begin{figure}
\begin{center}
\plotone{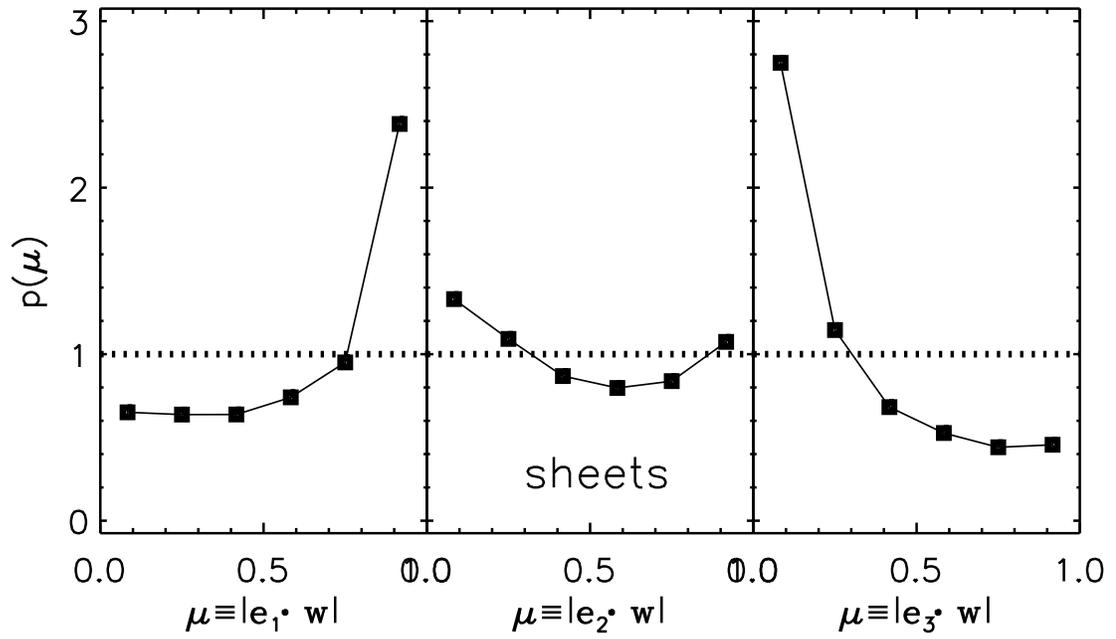}
\caption{Same as Figure \ref{fig:evor_knot} but for the sheet regions that the 
smallest shear eigenvalues are negative while the other two shear eigenvalues are 
positive.}
\label{fig:evor_fil}
\end{center}
\end{figure}

Figure \ref{fig:evor_void} plots $p(\mu)$ for the void regions, revealing that in the void regions the vorticity 
vectors are anti-aligned with the minor principal axes of the tidal tensors,  preferentially lying in the plane 
spanned by the major and intermediate principal axes, which is directly opposite to the case of the knot regions. 
This result is consistent with the numerical finding of \citet{libes-etal12b} even though in their work 
the vorticity field was filtered on much smaller scale.  
Figures \ref{fig:evor_fil} and \ref{fig:evor_pan} plot $p(\mu)$ for the filament and sheet cases, respectively. 
As can be seen, the filament regions show no strong signal of vorticity-shear alignments, while for the sheet 
case is found a clear signal alignments (anti-alignments) with the major (minor) principal axes but no 
strong alignments with the intermediate principal axes. 

\section{VORTICITY-SPIN ALIGNMENTS}\label{sec:spin}

The galaxy spin axes are hard to determine in practice, even under the simplified assumption that 
the minor axes of the galaxies are aligned with their spin axes. 
In the measurements of the alignments between the galaxy spin axes and the local vorticity vectors,  
the largest uncertainty would come from the inaccurate determination of the galaxy spin axes. Therefore, 
it is important to select carefully only those galaxies whose spin orientations can be determined with relatively  
high accuracy.  It is often assumed that for the case of the late-type spiral galaxies whose shapes 
are close to circular thin discs  their spin axes are orthogonal to the disc planes \citep{HG84}. 
Provided that  information on the position angles and axial ratios of the late-type spiral galaxies are available, 
their unit spin vectors can be determined up to the two-fold ambiguity in the sign of their radial components 
\citep{PLS00}.   The remaining uncertainty due to this two-fold ambiguity in the determination of the spin axes 
can be minimized by considering only those face-on or edge-on spirals.

As it is essential for our investigation to determine the directions of the galaxy angular momentum vectors as 
accurately as possible, we restrict our analysis only to the nearby large late-type spiral galaxies viewed either 
face on or edge on.  A sample of the nearby large late-type spiral galaxies was already obtained by \citet{lee11} 
from the SDSS DR7.  The sample contains those SDSS galaxies which have type Scd on the Hubble sequence 
\citep{huertas-etal11}, angular sizes larger than  $D_{\rm c}=7.92$ arcseconds in the redshift range of  
$0\le z\le 0.02$.  The value of this size cut-off $D_{\rm c}$ was imposed to remove the dwarfs which turned 
out to cause large uncertainties in the measurements of the spin axes due to their irregular shapes. 
Among the nearby large Scd galaxies in the sample of \citet{lee11},  we make a further selection of only those 
ones which have axial ratios larger than $0.9$ (nearly face-on) or smaller than $0.15$ (nearly edge-on) to 
minimize the uncertainties associated with the two-fold ambiguity. A total of $585$ nearby large late-type face-on 
(or edge-on) spirals are finally selected for our analysis. 

The unit spin vector of each selected galaxy, $\hat{\bf t}\equiv (\hat{t}_{x},\ \hat{t}_{y},\ \hat{t}_{z})$,  
is determined as \citep{lee11}
\begin{eqnarray}
\label{eqn:lx}
\hat{t}_{x}&=& \pm\cos\xi\sin(\pi/2-\delta)\cos\alpha + \vert\sin\xi\vert\sin P\cos(\pi/2-\delta)\cos\alpha  - 
\vert\sin\xi\vert\cos P\sin\alpha ,\\
\label{eqn:ly}
\hat{t}_{y} &=& \pm\cos\xi\sin(\pi/2-\delta)\sin\alpha + \vert\sin\xi\vert\sin P\cos(\pi/2-\delta)\sin\alpha + 
\vert\sin\xi\vert\cos P\cos\alpha ,\\
\label{eqn:lz}
\hat{t}_{z} &=& \pm\cos\xi\cos(\pi/2-\delta) - \vert\sin\xi\vert\sin P\sin(\pi/2-\delta) , 
\end{eqnarray}
where $P$ is the position angle of each selected galaxy, and $(\delta, \alpha)$ are the declination 
and right ascension of each galaxy's position vector expressed in the equatorial coordinate system, and 
the plus and minus signs in front of the first terms in Equation (\ref{eqn:lx})-(\ref{eqn:lz}) represents the 
two-fold ambiguity mentioned above. 
Here, $\xi$ is the galaxy's inclination angle related to its axial ratio $q$ and intrinsic flatness 
parameter $p$ as $\cos^{2}\xi = (q^{2}-p^{2})(1-p^{2})$. For the galaxies of type Scd, the intrinsic 
flatness parameter has the value of $p=0.1$ \citep{HG84}. 
Since the tidal shear and the vorticity fields from the 2MASS redshift survey have been reconstructed 
in the super-galactic coordinate systems,  we find a supergalactic expression for the unit spin vector of 
each selected galaxy, $\hat{\bf s}$, through the coordinate transformation of 
$\hat{\bf s} = R\hat{\bf t}$ where $R$ is the orthogonal matrix that transforms the equatorial  to the 
supergalactic frames.
 
By applying the Cloud-in-Cell interpolation (CIC) algorithm \citep{HE88} to the tidal shear fields reconstructed 
by \citet{LE07} from the 2MASS redshift survey,  we determine the tidal shear tensors at the positions of the 
selected SDSS galaxies and determine their principal axes, $\{{\bf e}_{1},\ {\bf e}_{2},\ {\bf e}_{e}\}$. 
Then, we calculate the cosines of the alignment angles, $\mu$, between the galaxy spin axes 
and the principal axes of the local tidal tensors as $\mu\equiv\vert{\bf s}\cdot{\bf e}_{i}\vert$.  
Recall that there are two different unit spin vectors assigned to each selected galaxy which differ from 
each other by the sign of the radial components.  As done in \citet{LE07},  we treat two spin vectors assigned 
to each galaxy as two independent realizations to end up having twice as many values of $\mu$ 
as the total number of the selected SDSS galaxies. 
Binning the values of $\mu$ and counting the number of those realizations, $n_{\mu}$, belonging to each 
$\mu$-bin,   we finally determine the probability density distribution, $p(\mu)$, of the spin-shear alignments,
calculating Poisson errors as $1/(n_{\mu}-1)^{1/2}$ associated with the determination of  $p(\mu)$.

\begin{figure}[ht]
\begin{center}
\plotone{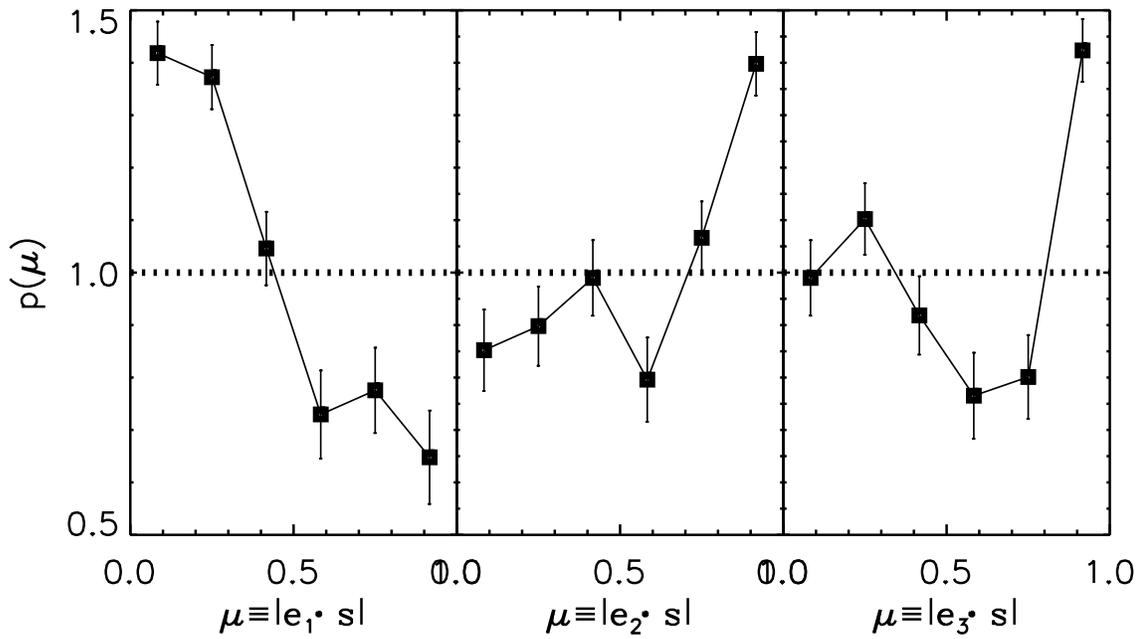}
\caption{Probability density distributions of the cosine of the angles between 
the the galaxy spin vectors and the major, intermediate and minor principal axes of the tidal 
shear tensors with Poisson errors in the left, middle and right panels, respectively.}
\label{fig:spinshear}
\end{center}
\end{figure}

Figure \ref{fig:spinshear} plots the probability density distributions of the cosines of the alignment angles 
between the unit spin vectors of the selected SDSS galaxies and the major, intermediate, and minor 
principal axes of the local tidal shear tensors with Poisson errors in the left, middle and right panels, 
respectively.  As can be seen, the spin axes of the selected galaxies seem to be strongly anti-aligned 
with the major principal axes of the local tidal tensors, but aligned with the intermediate and the minor 
principal axes. That is, the spin axes of the selected SDSS galaxies tend to lie in the plane perpendicular 
to the local direction of maximum matter compression.  

Compare our result with that of \citet{LE07} who found a significant signal of the correlations between the spin 
axes of the Tully galaxies and the intermediate principal axes of the local tidal tensors but no alignment signal 
with the other two principal axes.   We believe that the difference resulted from the inaccurate measurements 
of the spin axes of the Tully galaxies in the analysis of  \citet{LE07} which included not only the Scd galaxies 
but also the earlier type spirals with thick bulges  without taking into account the two-fold ambiguity. 
What is newly found from our analysis is that  the anti-alignments between the spin axes and the major principal 
axes are strongest while the alignments of the spin axes with the minor principal axes are as strong as 
that with the intermediate principal axes. Although the result is not completely against the tidal torque theory 
which predicts that the spin axes are preferentially aligned  with the intermediate principal axes, we would like 
to see if there exists strong spin-vorticity alignments which may help explain better the detected spin-shear 
alignment tendency.
 
\begin{figure}[ht]
\begin{center}
\plotone{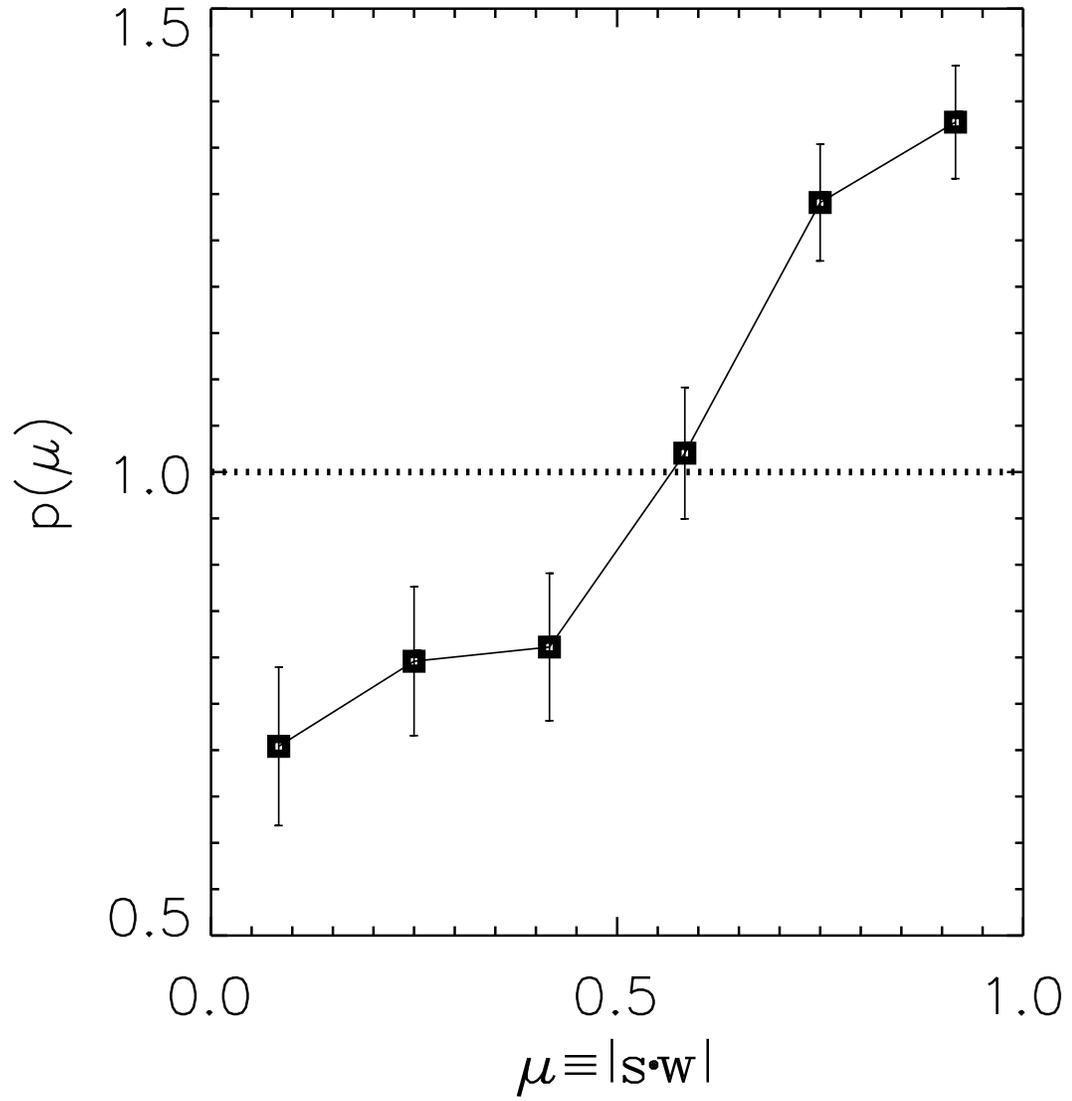}
\caption{Probability density distribution of the cosines of the angles between the 
vorticity vectors and the spin vectors of the nearby large face-on (or edge-on) Scd 
galaxies selected from the SDSS DR7 with Poisson errors.}
\label{fig:spinvor}
\end{center}
\end{figure}

Applying the CIC algorithm to the vorticity fields reconstructed in section \ref{sec:vor}, we calculate the local 
vorticity vectors at the positions of the selected SDSS disk galaxies. Then, we determine the probability 
distribution of the cosines of the angles between the unit spin vectors and the local vorticity vectors ${\bf w}$ 
in a similar manner, the result of which is plotted Figure \ref{fig:spinvor}. As can be seen, the probability density 
increases sharply and almost monotonically as $\mu$ increases, detecting a strong signal of the 
spin-vorticity alignments. We test the null hypothesis of no alignment (i.e., $p(\mu)=1$) 
with the help of the $\chi^{2}$-statistics and find that the null hypothesis is rejected at the $99.9999\%$ 
confidence level. 

The result shown in Figure \ref{fig:spinvor} is consistent with the numerical finding of \citet{libes-etal12b}, 
providing an observational support for the scenario that the galaxy angular momentum evolves via its 
interaction with the local vorticity in the nonlinear regime.  Recalling that in the work of 
\citet{libes-etal12b} the vorticity field was filtered on the galactic scale of $\le 1^{-1}\,h^{-1}$Mpc 
while in the current work the vorticity field is filtered on the much larger scale of $\sim 6\,h^{-1}$Mpc,   
we conclude that the vorticity effect wins over the tidal shear effect on the evolution of the galaxy angular 
momentum even in the quasi-nonlinear regime. 

\section{SUMMARY AND DISCUSSION}\label{sec:conclusion}

Utilizing the nearby large face-on (or edge-on) late-type spiral galaxies from the SDSS DR7 \citep{sdssdr7} 
and the tidal shear and vorticity fields reconstructed from the 2MASS redshift survey \citep{erdogdu-etal06}, 
we have measured the vorticity-shear, the spin-shear and the spin-vorticity alignments. The reconstructed 
vorticity fields have spatial resolution of $\sim 6\,h^{-1}$Mpc, corresponding to the quasi-nonlinear regime.
First, the vorticity vectors have been found to be strongly anti-aligned (aligned) with the major 
principal axes of the tidal shear tensors in the knot (void) regions, lying in the plane spanned by the other two 
principal axes.  Second, the galaxy spin axes have turned out to be strongly anti-aligned with the major principal 
axes of the local tidal shear tensors, while aligned with the intermediate and minor principal axes. 
Finally,  a clear signal of the spin-vorticity alignments has been detected, rejecting the null hypothesis of no spin-
vorticity alignment at $99.9999\%$ significance level. This result observationally stands by the new scenario  of 
\citet{libes-etal12b} that the tidally generated angular momentum of a galaxy subsequently evolves under the 
dominant effect of the vorticity field in the quasi-nonlinear and nonlinear regime.  

The spin-shear alignment tendency as well as its environmental dependence reported in the previous 
works can now be explained as follows. The cosmic flow in the nonlinear regime develops local vorticities 
which are preferentially inclined onto the plane orthogonal to the directions of either the maximum or the 
minimum volume compression depending on the web environment. The developed vorticities affect the galaxy angular 
momentum, modifying the spin-shear alignments from the initial tendency. 
Since it depends on the environment how fast the velocity field develops the vorticities and what 
directions the vorticity vectors get aligned with, the spin-shear alignments in the nonlinear regime come to 
take on the environmental dependence.  The stronger spin-shear alignments found in denser environments should 
result from the stronger vorticity effect there via which the nonlinear spin-shear alignments are established.
 
An interesting cosmological implication of our result is that the nonlinear vorticity fields and the the 
vorticity-induced galaxy alignments might be useful as a probe of cosmology and gravity.  
As shown in \citet{kitaura-etal12} and mentioned in \citet{libes-etal12b},  the vorticity of cosmic fluid  which 
equals zero in the linear regime grows in the nonlinear regime at third order.  The more rapidly a cosmic fluid 
develops the third order nonlinearity, the stronger the spin-vorticity alignments become due to the earlier onset of 
the vorticity effect on the evolution of the galaxy angular momentum.  In some modified gravity or dynamic dark 
energy models the high-order nonlinearity grows faster due to the presence of the fifth force  than for the 
$\Lambda$CDM case. Thus, in these alternative models the faster growth of the nonlinearity would lead 
to the more rapid development of the vorticity field, generating stronger spin-vorticity alignments. 
Very recently, \citet{li-etal13} claimed that the small scale powers of the divergence of the 
peculiar velocities should be a more sensitive probe of  modified gravity than the density power spectrum.
Given our result,  the small-scale powers of the absolute magnitude of the curl of the peculiar velocities might 
work powerfully as a complimentary probe.

\acknowledgments

I thank P.Erdogdu for providing me the data of the density and peculiar velocity fields from the 
2MASS redshift survey. I also thank M. Huertas-Company for providing information on the galaxy 
position angles and magnitudes.
This work was supported by the National Research Foundation of Korea (NRF)
grant funded by the Korea government (MEST, No.2012-0004195). Support for this work was also 
provided by the National Research Foundation of Korea to the Center for Galaxy Evolution 
Research (NO. 2010-0027910).
Funding for the SDSS and SDSS-II has been provided by the Alfred P. Sloan Foundation, 
the Participating Institutions, the National Science Foundation, the U.S. Department of 
Energy, the National Aeronautics and Space Administration, the Japanese Monbukagakusho, 
the Max Planck Society, and the Higher Education Funding Council for England. The 
SDSS Web Site is http://www.sdss.org/. 
The SDSS is managed by the Astrophysical Research Consortium for the Participating 
Institutions. The Participating Institutions are the American Museum of Natural History, 
Astrophysical Institute Potsdam, University of Basel, University of Cambridge, Case 
Western Reserve University, University of Chicago, Drexel University, Fermi lab, the 
Institute for Advanced Study, the Japan Participation Group, Johns Hopkins University, 
the Joint Institute for Nuclear Astrophysics, the Kavli Institute for Particle 
Astrophysics and Cosmology, the Korean Scientist Group, the Chinese Academy of Sciences 
(LAMOST), Los Alamos National Laboratory, the Max-Planck-Institute for Astronomy (MPIA), 
the Max-Planck-Institute for Astrophysics (MPA), New Mexico State University, Ohio State 
University, University of Pittsburgh, University of Portsmouth, Princeton University, 
the United States Naval Observatory, and the University of Washington.

\clearpage
\clearpage
\clearpage
\clearpage
\clearpage
\clearpage
\clearpage

\end{document}